# Full-dose Whole-body PET Synthesis from Low-dose PET Using High-efficiency Denoising Diffusion Probabilistic Model: PET Consistency Model

Shaoyan Pan[1,2], Elham Abouei[1], Junbo Peng[1], Joshua Qian[1], Jacob F Wynne[1], Tonghe Wang[3], Chih-Wei Chang[1], Justin Roper[1], Jonathon A Nye[4], Hui Mao[5] and Xiaofeng Yang[1,2*]

[1]Department of Radiation Oncology and Winship Cancer Institute, Emory University, Atlanta, GA 30322, USA

[2]Department of Biomedical Informatics, Emory University, Atlanta, GA 30322, USA

[3]Department of Medical Physics, Memorial Sloan Kettering Cancer Center, New York, NY 10065, USA

[4]Radiology and Radiological Science, Medical University of South Carolina, Charleston, SC 29425, USA

[5]Department of Radiology and Imaging Science, and Winship Cancer Institute, Emory University, Atlanta, GA 30322, USA

*Email: xiaofeng.yang@emory.edu

**Abstract**

**Objective:** Positron Emission Tomography (PET) has been a commonly used imaging modality in broad clinical applications. One of the most important tradeoffs in PET imaging is between image quality and radiation dose: high image quality comes with high radiation exposure. Improving image quality is desirable for all clinical applications while minimizing radiation exposure is needed to reduce risk to patients.

**Approach:** We introduce PET Consistency Model (PET-CM), an efficient diffusion-based method for generating high-quality full-dose PET images from low-dose PET images. It employs a two-step process, adding Gaussian noise to full-dose PET images in the forward diffusion, and then denoising them using a PET Shifted-window Vision Transformer (PET-VIT) network in the reverse diffusion. The PET-VIT network learns a consistency function that enables direct denoising of Gaussian noise into clean full-dose PET images. PET-CM achieves state-of-the-art image quality while requiring significantly less computation time than other methods. Evaluation with normalized mean absolute error (NMAE), peak signal-to-noise ratio (PSNR), multi-scale structural similarity index (MS-SSIM), normalized cross-correlation (NCC), and clinical evaluation including Human Ranking Score (HRS) and Standardized Uptake Value (SUV) Error analysis shows its superiority in synthesizing full-dose PET images from low-dose inputs.

**Results:** In experiments comparing eighth-dose to full-dose images, PET-CM demonstrated impressive performance with NMAE of 1.278±0.122%, PSNR of 33.783±0.824dB, SSIM of 0.964±0.009, NCC of 0.968±0.011, HRS of 4.543, and SUV Error of 0.255±0.318%, with an average generation time of 62 seconds per patient. This is a significant improvement compared to the state-of-the-art diffusion-based model with PET-CM reaching this result 12x faster. Similarly, in the quarter-dose to full-dose image experiments, PET-CM delivered competitive outcomes, achieving an NMAE of 0.973±0.066%, PSNR of 36.172±0.801dB, SSIM of 0.984±0.004, NCC of 0.990±0.005, HRS of 4.428, and SUV Error of 0.151±0.192% using the same generation process, which underlining its high quantitative and clinical precision in both denoising scenario.

**Significant:** We propose PET-CM, the first efficient diffusion-model-based method, for estimating full-dose PET images from low-dose images. PET-CM provides comparable quality to the state-of-the-art diffusion model with higher efficiency. By utilizing this approach, it becomes possible to maintain high-quality PET images suitable for clinical use while mitigating the risks associated with radiation.

# I. Introduction

Positron Emission Tomography (PET) has broad clinical applications for diagnosis, prognosis, and treatment planning for oncology [1], cardiology [2], and neurology [3]. PET images can be used to determine the disease stage, lesion malignancy, and response to treatment [4], and the images can also be co-registered with Computed Tomography (CT) or magnetic resonance imaging (MRI) scans to visualize both anatomic and metabolic localization [5]. One of the most important tradeoffs in PET imaging is between image quality and radiation dose: high image quality comes with high radiation exposure. Improving image quality is desirable for all clinical applications while minimizing radiation exposure is needed to reduce risk to patients.

Accordingly, various techniques (mainly machine learning techniques) have been developed to minimize the doses of PET scans, while preserving high image quality. These methods include multilevel canonical correlation analysis [6], mapping-based sparsed representation [7], and semisupervised tripled dictionary learning [8]. These advances allow reductions of administered radiotracer or scan time required for obtaining full-dose PET images [9]. Nonetheless, the performances of these approaches depend on handcrafted features extracted based on prior domain knowledge, which is incomprehensive to describe all the anatomical structure and texture details of the PET images, limiting the quality and visual reality of the denoising PET images.

Recently, deep learning approaches [4,5,10,11] using deep auto-context convolutional neural networks (CNN) [12] have been introduced for PET image denoising. State-of-the-art methods including three-dimensional conditional generative adversarial networks (CGAN) [8] and cycle-consistent GAN [13], deploy adversarial training strategy to enable the denoised PET image to be not only with high accuracy compared to the full dose PET image but also with highly realistic visual appearance [14-17]. These GAN-based models employ a dual-network architecture that operates in an adversarial manner: a generator tasked with converting low-dose PET images into full-dose equivalents, and a discriminator that ensures the generator's outputs closely align with the statistical distribution of ground-truth full-dose PET images. The optimization function of the discriminator are equivalent to minimizing the Jensen-Shannon divergence [18] between the distribution of the synthetically generated full-dose PET images and the ground-truth, resulting in outputs with a highly realistic visual appearance. However, the GAN-based models reveal high instability in the training process due to the adversarial strategy, which also limits the accuracy of the structure details of the PET denoising. Accordingly, Denoising Diffusion Probabilistic Models (DDPM) and Score-matching Models present an alternative generative approach. DDPM methods employ a Markov chain of diffusion steps to systematically introduce random noise into data and subsequently learn to reverse the diffusion process iteratively, generating desired data samples from the noise [19,20]. They have been proven to be advantageous over GANs, in both natural image [21-27] and medical image synthesis in other different modalities [28-32], with improved training stability and the production of more authentic and higher-quality synthetic images [29]. However, its iterative reverse process requires much more time to generate images. Accordingly, Song *et al.* proposed a Consistency Model (CM), which can generate images with quality comparable to DDPMs, but with much higher efficiency [33]. Instead of a step-by-step reversal, iterating backwards through each noise addition stage as DDPMs in the reverse process, the CM is designed to learn a function that can directly map any noisy image, denoted as $X_t$, back to its clean, original form $X_0$, irrespective of the noise level '$t$'. As a result, the CM can regenerate high-quality images in just a few iterations, substantially cutting down the generation time.

In this study, we propose PET-CM to generate full-dose PET images from low-dose images. It is the first CM-based model for conditional medical image synthesis and also for PET denoising. Drawing upon the innovative integration of the Shifted-window Vision Transformer V-Net (called PET-VIT) into the consistency model, the PET-CM is able to capture both global and local features effectively, forging a pathway to richer and more accurate representations of PET images therefore promising a significant improvement in both the efficiency and quality of the synthetic PET images. Furthermore, our strategy adds a new loss regularization term to the CM process to minimize the difference between the truth and generated full-dose PET images. This term can benefit the consistency model in two primary aspects: stabilizing the training phase fundamentally and augmenting the pixel-level detail's accuracy of the synthetic images. The proposed PET-CM mitigates the well-acknowledged training instabilities that have been a notable limitation in existing GAN-based models, and requires much less time to generate synthetic images compared to the diffusion-based models. The PET-CM was evaluated through comparative analyses against contemporaneous methods like PET-CGAN, PET-cycleGAN, PET-DDPM, and PET-IDDPM utilizing eighth- and quarter-dose PET images. Our quantitative evaluation resoundingly demonstrates that the PET-CM can synthesize full-dose PET images with state-of-the-art accuracy, surpassing competing methodologies in either quality and efficiency, thereby evidencing a pioneering stride towards redefined boundaries in PET image denoising.

## II. Method

The proposed PET-CM framework includes a two-step process: the forward process introduce noise into the data $X$ through a controlled process including multiple timestep $t$; the reverse process trains a deep learning model to denoise the image noise from the forward process to create high-quality synthetic images from pure noise samples, called the reverse process. While most conventional diffusion models consider the reverse process an iterative Markov process so we can generate $X_{t-1}$ from $X_t$ at each timestep $t$, enabling the recursive generation of $X_0$ through multiple timesteps; PET-CM trained a neural network, namely PET Shifted-window vision transformer (PET-VIT) that directly connects $X_T$ to $X_0$. This direct connection allows for the generation of $X_0$ in just a few steps, bypassing the need for extensive recursion so greatly reduce the generation time, as shown in Fig. 1 a). In addition, a low-dose PET scan is concatenated to the input noisy image of the neural network through the whole reverse process, to generate paired full-dose PET for the same patients.

Figure 1.b) illustrates the design of PET-VIT, which is based on Pan *et al.*'s [29] denoiser architecture. The architecture comprises a two-dimensional encoder and decoder. The encoder includes down-sampling residual convolutional blocks and Shifted-window self-attention (we called it Swin transformer) blocks [34], which enable the learning of semantic features at various resolution levels. Conversely, the decoder exhibits a symmetrical structure to the encoder and is responsible for reconstructing the final estimations using the learned semantic features. We show the details of network architecture in Appendix. A. Below, we provide a detailed description of the mathematical formulation for PET-CM.

a) PET-CM diffusion pipeline

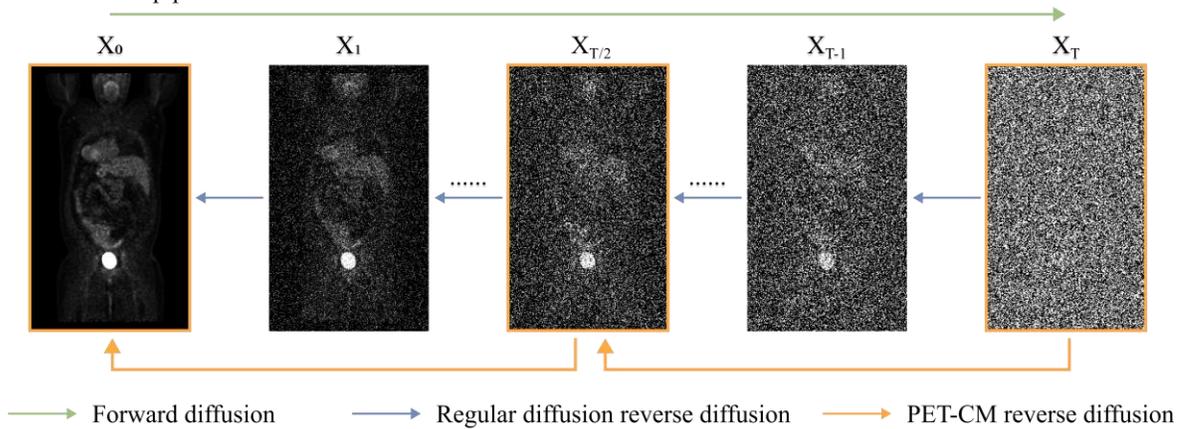

b) PET-VIT architecture

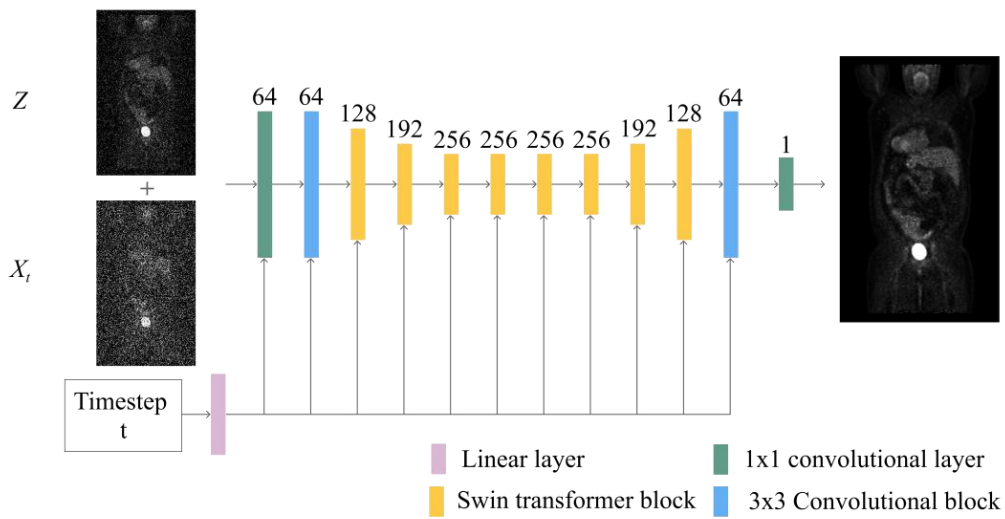

**Figure 1:** a) The denoising diffusion process of PET-CM employs a forward diffusion process to introduce Gaussian noise to full-dose PETs using multiple timesteps, gradually transforming them into pure Gaussian noise. Contrast to the conventional reverse diffusion, PET-CM generates clean images with larger intervals between timesteps, providing efficient denoising without requiring a large number of iterations. b) Network architectures of PET-VIT used in PET-CM: A symmetrical encoder-decoder architecture is employed to learn the reverse process, in which PET-CM learns a consistency function for generating full dose images.

**II.A PET Consistency model**

A standard conditional consistency model consists of three processes: First, a forward process gradually applies small amounts of Gaussian noise $N$ to an initial full dose PET image $X_0$ over a series of $T$ timesteps. This transforms the PET scan into pure multi-dimensional Gaussian noise $X_T$. Following this, our developed PET-VIT is configured to learn a reverse diffusion process, considering the supplementary low-dose PET scan (Z) as a guiding variable. Distinct from established diffusion models which progressively eliminate the incremental noise incorporated at each interval, reverting $X_T$ to its pristine state of $X_0$, our consistency model envisages a more efficient trajectory. It orchestrates the PET-VIT to identify a pathway that encompasses the viable solutions for all timesteps, facilitating a non-recursive derivation of $X_0$ directly from $X_T$. In the third process, by employing an optimal PET-CM denoiser, we can run a sampling process to recursively translate a pure Gaussian noise to a full dose PET image corresponding to the input low dose PET.

II.A.1 Forward process of the Consistency Model

During the forward process in, we aim to generate noisy images $[X_1, X_2, ..., X_T]$ so we can have data to train the PET-VIT for denoising. Specifically, the process of generating the noisy image can be described as a Markov process, where the transition probability from image $X_{t-1}$ to image $X_t$ follows independent Gaussian distribution: $q(X_t | X_{t-1}) = \mathcal{N}(X_t; \sqrt{1-\beta_t} X_{t-1}, \beta_t I)$ with a pre-determined variance $\beta_t$. Assuming a large timestep T, we can generalize the forward diffusion process into continuous-time processes. We follow Song *et al.*'s work [20] to develop the forward process in a form of stochastic differential equations (SDE):

$$dX_t = f(X_t, t)dt + g(t)dw_t \quad [1]$$

where $dw_t$ is a normal noise (the derivative of a Wiener process $w_t$ is a normal noise).

In implementation, we need to build a set of timesteps $t = [t_1, t_2, ..., t_J]$ suitable for the consistency model:

$$t_j = \left(\varepsilon^{\frac{1}{\rho}} + \frac{j-1}{J-1}\left(T^{\frac{1}{\rho}} - \varepsilon^{\frac{1}{\rho}}\right)\right)^{\rho} \quad [2]$$

where we empirically set the minimum noise scale $\varepsilon = 0.001$, the maximum noise scale $T = 100$, the number of timesteps $J = 150$, and noise power $\rho = 7$. Then, we can generate the noisy full dose PET $X_{t_j}$ at timestep $t_j$:

$$X_{t_j} = X_0 + t_j \epsilon_{t_j} \quad [3]$$

where $\epsilon_{t_j} \sim \mathcal{N}(0, I)$ is a random normal noise.

II.A.2 Reverse process of the Consistency Model

Based on Eqn. 1, we have a Probability Flow ordinary differential equation (PF-ODE) reverse process as:

$$dX_t = \left[f(X_t, t) - \frac{1}{2}g^2(t)\nabla \log p_t(X_t)\right] dt \quad [4]$$

By setting $f(X_t, t) = 0$ and $g(t) = \sqrt{2t}$, we arrive at:

$$\frac{dX_t}{dt} = -tS_\theta(X_t, t) \quad [5]$$

where $S_\theta(X_t, t)$ is the estimation of the score function $\nabla \log p_t(X_t)$ and the PF-ODE can be solved by any numerical ODE solver, such as Heun's method [33] (details shown in Appendix. B).

Obtaining the clean image $X_0$ using ODE solvers requires iterative evaluations of the score model $S_\theta(X_t, t)$, which is a computationally costly process. To accelerate sampling, the self-consistency property is introduced to arbitrary pairs of $(X_t, t)$, under the guidance of $Z$, in the same PF-ODE trajectory, which is formulated as:

$$c_\theta(X_t, t, Z) = c_\theta(X_\epsilon, \epsilon, Z) = X_\epsilon \approx X_0, \ t \in [\epsilon, T] \quad [7]$$

where $c_\theta$ is the consistency function learned by the PET-VIT $\theta$, mapping all the points in the trajectory to the destination of the trajectory $X_0$, and the consistency function is estimated by a neural network model $\theta$. $\epsilon$ is a scalar sufficiently close to 0. We can thus translate a pure Gaussian noise $X_T$ into a full dose PET $X_0$ conditioning on a low-dose PET $Z$ and, in an extreme case, the full-dose PET can be generated by one step as $X_0 \approx c_\theta(X_T, T, Z))$.

A straightforward way to train the consistency model $c_\theta$ is to perform distillation of the pre-trained score function $S_\theta$, which is called the consistency distillation (CD) method. However, the CD strategy requires pre-trained model training on an extremely large-scale dataset, which is impractical in medical image synthesis. Alternatively, we can directly train the consistency model without knowledge of the score model $S_\theta$ in what is called the consistency training (CT) method. In the CT strategy, we avoid the pre-trained score model by leveraging the unbiased estimator [33]

$$\nabla \log p_t(X_t, Z) = -E\left[\frac{X_t - X_0}{t^2} | X_t, Z\right] \quad [8]$$

and the PET-VIT $\theta$ is optimized by:

$$L = MAE\left(c_\theta(X_{t_j}, t_j, Z), c_\theta(X_{t_{j-1}}, t_{j-1}, Z)\right) \quad [9]$$

where $t_j$ and $t_{j-1}$ are adjacent time steps for the same PF-ODE trajectory, and $X_{t_{j-1}}$ is calculated by Heun's method from the $X_{t_j}$. The loss function is designed to minimize the output of two adjacent data points

$(X_{t_{j-1}}, t_{j-1})$ and $(X_{t_j}, t_j)$. When the subscript $j$ runs through all possible steps, the consistency property of $c_\theta$ is enforced on all the data points in the same trajectory.

Furthermore, to address denoising tasks more proficiently, we introduce a regularization term within the loss function. This addition is designed to steer the synthesis of full-dose PET images towards a configuration where pixel-level discrepancies with the ground truth are substantially minimized, ensuring higher fidelity in the generated output:

$$L_p = MAE\left(c_\theta(X_{t_j}, t_j, Z), X_0\right) \quad [10]$$

The final objective function can be expressed as:

$$L_{total} = L + \gamma L_p \quad [11]$$

where $\gamma$ is empirically chosen as 0.5.

Following Song [33], we employed the exponential moving average (EMA) technique to boost the convergence of the optimization problem. Specifically, the optimization process requires two identical networks: $\theta$ and $\theta'$. The loss function is minimized by stochastic gradient descent on the model parameter set $\theta$, while updating $\theta'$ with the EMA of $\theta$. The modified loss function can be expressed as:

$$L_{total} = MAE\left(c_\theta(X_{t_j}, t_j, Z), c_{\theta'}(X_{t_{j-1}}, t_{j-1}, Z)\right) + \gamma L\_p \quad [12]$$

II.A.3 Consistency Sampling

With a well-trained consistency denoiser $\theta$, we can generate a full-dose PET image at timestep $t_{j'}$ at any arbitrary timestep $t_j$:

$$X_{t_{j'}} = c_\theta(X_{t_j}, t_j, Z) + \sqrt{t_{j'}^2 - \varepsilon^2}\epsilon_{t_j} \quad [13]$$

In our consistency model experiments, we adopt two-step generation using the set of timesteps $t = [1, 75, 150]$ for inference and $X_{t_1}$ is the final synthetic full dose PET images (sPET). Utilizing timesteps $[1, 75, 150]$, the process initiates with $X_{t_j}$ at $t_j = 150$, representing pure Gaussian noise, to compute a less noisy image $X_{t_j'}$ at $t_j' = 75$ as Eqn. 13. This yields an intermediate noisy PET image, denoted as $X_{75}$, which, when applied with the same equation, for the calculation of $X_1$, the final sPET. This timestep is accommodated on the initial application of 150 timesteps during the training phase, which introduce the maximal level of noise. To account for the randomness introduced during the generation process, each full-dose image is generated three times, and the final output is obtained by averaging the results. This approach follows a Monte Carlo-based (MC-based) generation method, which helps to mitigate the impact of random noise and improve the overall stability and reliability of the generated PETs [28].

## III. Data acquisition

### III.A Institutional whole-body PET dataset

The dataset consists of 35 patients (20 with cancer at various sites) contain 11,200 slices from whole-body $^{13}$F-FDG PET/CT with a pixel size of 3.65 x 3.65 x 3.27 mm³. A five-fold cross-validation methodology was adopted for the study. In this schema, four out of five subsets—each comprising 28 patients, which collectively contribute 8,960 image slices—were allocated for model training during each fold. Simultaneously, the fifth subset, consisting of 13 patients and totaling 2,240 slices, was sequestered for model validation. This process was iteratively conducted, with the model undergoing training and assessment across the differing folds, ensuring that each fold was evaluated. Images were acquired using a GE Discovery 690 PET/CT scanner. For patients with BMI less than or greater than 30, 370 MBq and 440 MBq $^{18}$F-FDG radiotracer was administrated, respectively, followed by 60-minute uptake period. All subjects were scanned with 2 bed positions at 2.5 minutes. A 3D ordered-subset expectation maximization (OSEM) algorithm of 3 iterations and 24 subsets was used for image reconstruction, and a CT-based attenuation correction was applied. To create low-dose PET data, two additional sets of PET data were histogram-reduced to one eighth and one quarter of the original bed duration for all bed positions.

All full- and low-dose PETs were centered and cropped to a size of 96 x 192 x 320. PET intensities were then jointly normalized to [-1, 1]. Networks were trained and tested on 2D axial slices. These slices were subsequently combined to reconstruct a complete whole-body scan for evaluation. For the purpose of reconstructing a complete whole-body scan for evaluation, we meticulously combined individual slices. Throughout both the low-dose dataset preparation and the training phase, we consistently employed a patch size of 64x64. During the inference phase, a sliding window prediction strategy was utilized to segment the boundary-padded PET axial slices, which were expanded to dimensions of 128x192, into numerous patches of identical size to those used in training. Each patch was independently processed by the PET-CM to predict the corresponding full-dose PET patches. These were then meticulously assembled to reconstruct the complete PET image, which was boundary-cropped to restore its original size of 96x192, thus yielding our final output. During training, we utilized an AdamW optimizer with an initial learning rate of 2 x 10$^{-5}$ and a weight decay of 1 x 10$^{-4}$ across 500 epochs to train PET-CM. In the PET-CM experiments, the model was trained with 150 timesteps and generated full dose PETs with 2 steps ($t = [1, 75, 150]$).

## IV. Implementation and performance evaluation

### IV.A Implementation details

The PET-CM frameworks and competing networks were executed on a workstation running Microsoft Windows 11 equipped with a single NVIDIA RTX 6000 GPU with 48 gigabytes of memory. The experiments were conducted using the PyTorch framework in Python version 3.8.11.

IV.B Quantitative evaluation

The quality of the synthetic PET generated by the proposed PET-CM were evaluated using the normalized mean absolute error (NMAE in percentage of the activity concentration value per volume), peak signal-to-noise ratio (PSNR in decibel (dB)), multi-scale structure similarity index (MS-SSIM, with an evaluation scale of 3) and maximum normalized cross correlation (NCC) indices. NMAE is a negatively-oriented score: a smaller value indicates a smaller absolute difference between the ground truth PET and the sPET images, implying better sPET quality. PSNR, SSIM and NCC are positively-oriented: higher PSNR indicates a higher peak signal similarity, higher SSIM indicates greater overall visual similarity, and higher NCC indicates better image correlation between the sPET and PET images. All metrics were background excluded and calculated at the patient level. Furthermore, our study incorporates the expertise of radiation oncologist and medical physicist for clinical evaluation purposes. Initially, sPET images generated by the PET-CM and four other distinct methodologies, alongside the ground truth full-dose PET images, were independently assessed by the two clinicians on a scale from 1 to 6, with higher scores indicating superior sPET image quality. Following this, abnormal regions present in all patients were identified. In these identified areas, we computed the Standardized Uptake Value (SUV) Error in percentage terms. This measure is based on the absolute difference in SUVmean between the sPET's tumor region and the ground truth PET's corresponding area, adjusted by the highest intensity values in the ground truth PET. The evaluations culminated in an average value of all patients for each metric. To establish a benchmark, we compared the proposed networks' performance against state-of-the-art low-dose to full-dose PET conversion methods, including 3D PET-CGAN [13], 3D PET-cycleGAN [13], 2D PET-DDPM [19], and 2D PET improved DDPM (PET-IDDPM) [35]. We configured the GAN-based networks as detailed in their corresponding references. For PET-DDPM and PET-IDDPM, we adopted the same network architecture from PET-CM for equipoise. We applied identical training hyperparameters for all networks, including training and generation timesteps (applicable only for diffusion-based methods), and data preprocessing. Furthermore, all employed methods rigorously followed a Monte Carlo (MC)-based generation strategy. This entailed conducting three separate runs for each method, subsequently averaging the outcomes of these runs to yield the result. We conducted pair-wise comparisons using one-sampled Student's paired t-test with α = 0.05.

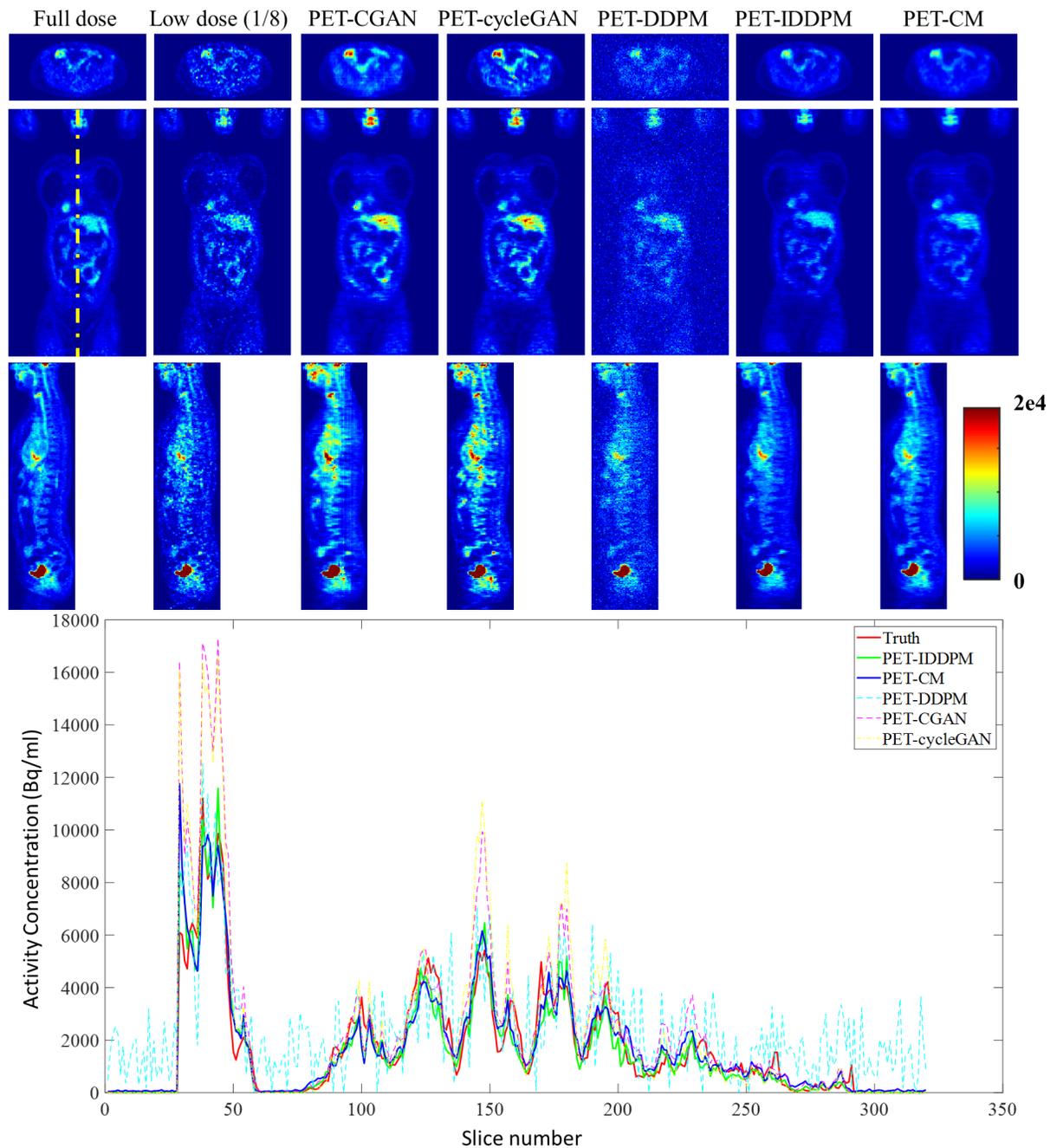

**Figure 2**: Visualization of the imaging results and single line profiles for the synthetic images obtained from eighth dose PETs. The figure displays the full dose image (1st column), the corresponding low dose image (2nd column), and the sPET images generated by different methods: PET-CGAN (3rd column), PET-cycleGAN (4th column), PET-DDPM (5th column), PET-IDDPM (6th column), and PET-CM (7th column). The single-line profiles, represented by the yellow line traversing the axial slices of the full-dose PET, are displayed beneath the visual representations. These profiles depict the fluctuation of intensity values along the line in both the authentic and synthesized images. The color red is used to denote the ground truth full-dose PET images.

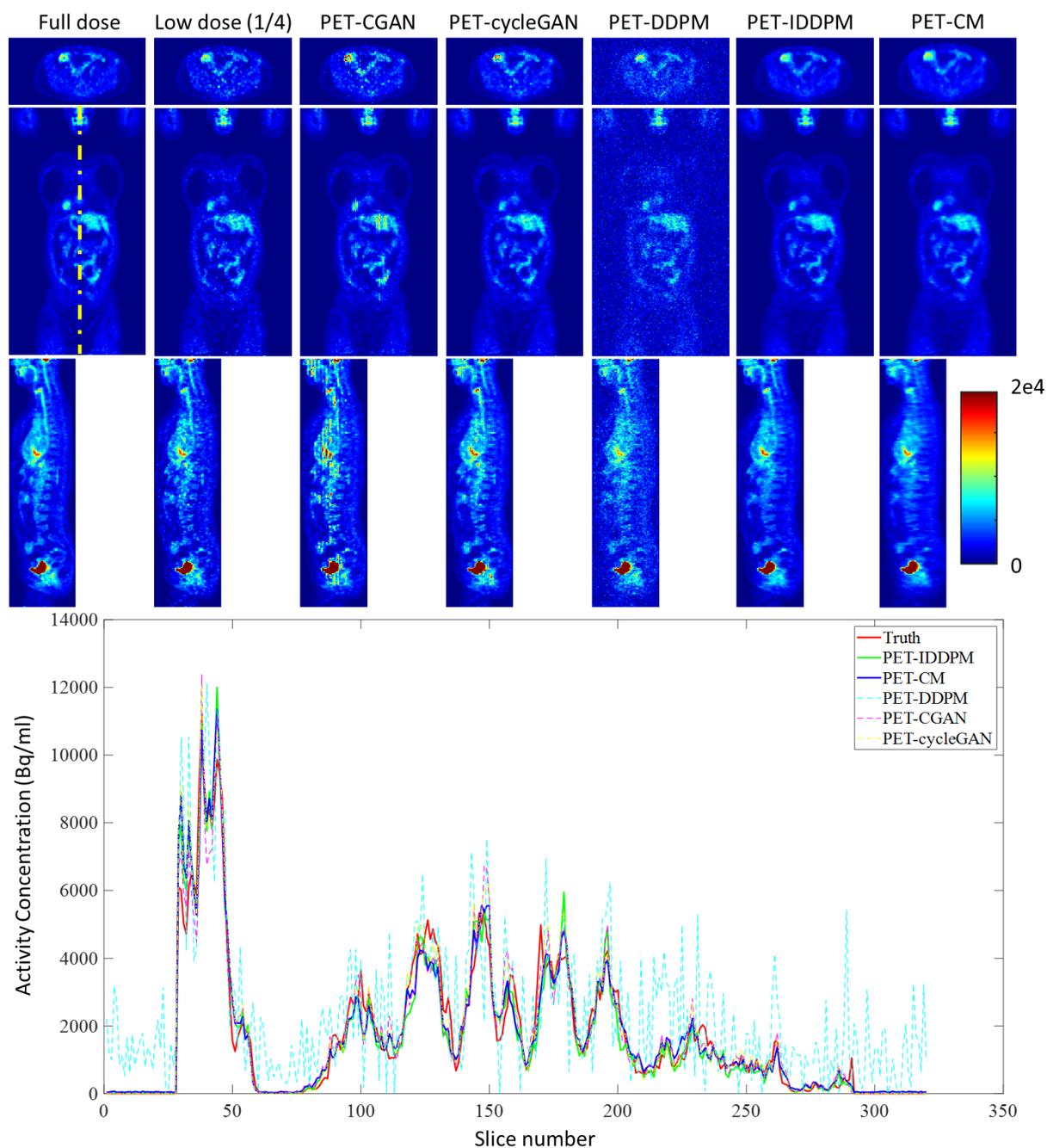

**Figure 3**: Visualization of the imaging results and single line profiles for the synthetic images obtained from quarter-dose PETs. The figure displays the full dose image (1st column), the corresponding low dose image (2nd column), and the sPET images generated by different methods: PET-CGAN (3rd column), PET-cycleGAN (4th column), PET-DDPM (5th column), PET-IDDPM (6th column), and PET-CM (7th column). The single-line profiles, represented by the yellow line traversing the axial slices of the full-dose PET, are displayed beneath the visual representations. These profiles depict the fluctuation of intensity values along the line in both the authentic and synthesized images. The color

### IV.C Quantitative study for network selection and generation timesteps

To comprehensively evaluate the performance of our proposed method and determine the optimal conditions for each approach, we conducted several studies. The first study aimed to assess the impact of different state-of-the-art deep learning neural networks on the quality of synthetic full dose PET images. We compared the performance of three networks: Swin Vision Transformer (PET-VIT, proposed), U-net [36], and token-based multi-layer linear Mixer Unet (MLP-Unet) [37]. Unet is constructed by replacing Swin transformer blocks with convolutional blocks (shown in Fig. 1), and MLP-Unet is constructed using 2D token-based MLP-Mixer blocks instead of Swin transformer blocks.

In the second study, we examined the effect of generation timesteps on the quality of synthetic PET images. We varied the generation timestep for PET-CM using even spaced timesteps ranging from 1 to 150, with step sizes 1, 2 (proposed), 5, and 10. These studies provide valuable insights into the optimal neural network architecture and generation timestep for achieving high-quality synthetic PET images.

### V. Result

Synthetic full-dose PETs generated from low-dose images using PET-CM and competing networks are presented in Fig. 2 and 3, accompanied by single-line image profiles. In the comparative analysis of sPET images, it is observed that those generated through PET-CM and PET-IDDPM exhibit a significantly higher degree of realism in comparison to those produced by PET-CGAN and PET-cycleGAN, in the context of synthesizing full-dose PET images from eighth- and quarter-dose PET images. The line profiles, illustrated beneath the visual representations, further corroborate the superiority of PET-CM and PET-IDDPM. These profiles demonstrate that the activity concentration delineated along the designated yellow line, traversing various axial sections, bears a closer visual resemblance to the ground truth full-dose PET images. To evaluate the quality of the synthetic images, we conducted quantitative assessments, the results of which are presented in Table. 1 for the eighth-dose PETs and Table. 2 for the quarter-dose PETs. Additionally, we performed additional experiments to analyze the impact of network selection. Results are summarized in Table. 3. Results of varying generation timesteps for PET-CM are given in Table. 4.

**Table 1**: Quantitative analysis of synthetic full-dose images from PET-CM vs. PET-CGAN, PET-cycleGAN, PET-DDPM, and PET-IDDPM using the eighth-dose PETs. (↑) indicates positive-orientation: greater values indicate better performance; (↓) indicates negative orientation: smaller values indicate better performance. The generation time (G-time) for each patient is reported. The best-performing network(s) are highlighted in bold, and the second-best network(s) are underlined, based on mean evaluation results. P-values between the competing networks and PET-CM

are shown below each method. Values are rounded to three decimals.

| 1/8 dose to full dose | NMAE (%) (↓) | PSNR (dB) (↑) | SSIM (↑) | NCC (↑) | HRS (↑) | SUV Error (%) (↓) | G-time (Sec) |
|---|---|---|---|---|---|---|---|
| PET-CM | <u>1.278±0.122</u> | **33.783±0.824** | **0.964±0.009** | **0.968±0.011** | **4.543** | <u>0.255±0.318</u> | 62.321 |
| *p-value* (PET-CM) | N/A | N/A | N/A | N/A | N/A | N/A | |
| PET-IDDPM | **1.211±0.104** | <u>33.217±0.798</u> | <u>0.960±0.011</u> | <u>0.966±0.012</u> | <u>4.457</u> | **0.253±0.376** | 823.713 |
| *p-value* (PET-CM) | 0.728 | 0.711 | 0.462 | 0.931 | N/A | 0.978 | |
| PET-DDPM | 3.871±0.322 | 23.281±0.298 | 0.823±0.046 | 0.852±0.049 | 1.000 | 1.082±0.761 | 817.326 |
| *p-value* (PET-CM) | <0.010 | <0.010 | <0.010 | <0.010 | N/A | <0.010 | |
| PET-CGAN | 2.521±0.358 | 26.782±2.381 | 0.936±0.020 | 0.934±0.020 | 2.614 | 0.547±0.640 | **32.142** |
| *p-value* (PET-CM) | <0.010 | <0.010 | <0.010 | <0.010 | N/A | <0.010 | |
| PET-cycleGAN | 2.212±0.274 | 27.180±1.702 | 0.941±0.016 | 0.939±0.017 | 3.100 | 0.555±0.639 | **32.142** |
| *p-value* (PET-CM) | <0.010 | <0.010 | <0.010 | <0.010 | N/A | <0.010 | |

**Table 2**: Quantitative analysis of synthetic full-dose images from PET-CM vs. PET-CGAN, PET-cycleGAN, PET-DDPM, and PET-IDDPM using the quarter-dose PETs. (↑) indicates positive-orientation: greater values indicate better performance; (↓) indicates negative orientation: smaller values indicate better performance. The generation time (G-time) for each patient is reported. The best-performing network(s) are highlighted in bold, and the second-best network(s) are underlined, based on mean evaluation results. P-values between the competing networks and PET-CM are shown below each method. Values are rounded to three decimals.

| 1/4 dose to full dose | NMAE (%) (↓) | PSNR (dB) (↑) | SSIM (↑) | NCC (↑) | HRS (↑) | SUV Error (%) (↓) | G-time (Sec) |
|---|---|---|---|---|---|---|---|
| PET-CM | <u>0.973±0.066</u> | <u>36.172±0.801</u> | **0.984±0.004** | **0.990±0.005** | <u>4.428</u> | **0.151±0.192** | 62.321 |
| *p-value* (PET-CM) | N/A | N/A | N/A | N/A | N/A | N/A | |
| PET-IDDPM | **0.956±0.054** | **36.723±0.681** | **0.984±0.003** | <u>0.987±0.004</u> | **4.571** | <u>0.155±0.201</u> | 823.713 |
| *p-value* (PET-CM) | 0.826 | 0.584 | 0.963 | 0.810 | N/A | 0.847 | |
| PET-DDPM | 3.738±0.398 | 24.114±0.238 | 0.831±0.074 | 0.857±0.050 | 1.000 | 1.020±0.692 | 817.326 |
| *p-value* (PET-CM) | <0.010 | <0.010 | <0.010 | <0.010 | N/A | <0.010 | |
| PET-CGAN | 1.762±0.200 | 30.125±1.371 | 0.948±0.012 | 0.940±0.015 | 2.557 | 0.469±0.481 | **32.142** |
| *p-value* (PET-CM) | <0.010 | <0.010 | <0.010 | <0.010 | N/A | <0.010 | |
| PET-cycleGAN | 1.382±0.182 | 31.712±0.752 | 0.955±0.010 | 0.955±0.022 | 3.157 | 0.423±0.440 | **32.142** |

| | | | | | | |
|---|---|---|---|---|---|---|
| *p-value* (PET-CM) | <0.010 | <0.010 | <0.010 | <0.010 | N/A | <0.010 |

### V.1 Comparison of eighth-dose to full-dose PETs

The PET-CM secured the second-best results, recording an NMAE of 1.278±0.122% and SUV Error of 0.255±0.318%. It excelled by achieving the highest SSIM of 0.964±0.009, a PSNR of 33.783±0.824dB, an NCC of 0.968±0.011, and HRS of 4.543. Statistical analysis confirms that there is no significant difference between the PET-CM and PET-IDDPM. The PET-CM demonstrates significantly faster generation time than the PET-IDDPM (62 seconds vs. 823 seconds to generate one patient's image), though both proposed methods are slower than GAN-based methods. Comparatively, the proposed PET-CM exhibits substantial quantitative improvements at all image-quality metrics with statistical significance ($p<0.05$) compared to other methods.

### V.2 Quantitative result of quarter-dose to full-dose PETs

The PET-CM exhibited superior performance, achieving the highest SSIM at 0.984±0.004, an NCC of 0.990±0.005, and a SUV Error of 0.151±0.192%. It also recorded the second-best results with an NMAE of 0.973±0.066% a PSNR of 36.172±0.801dB, and a HRS of 4.428, closely matching the performance attained by the PET-IDDPM. Compared to the PET-IDDPM, the PET-CM does not show any statistical difference. However, when compared to other methods, the PET-CM shows significant quantitative and statistical improvements (all $p < 0.05$). This finding highlights the superiority of diffusion-based models over the commonly used GAN-based models in accurately estimating full-dose images from low-dose inputs.

### V.3 Network selection and consistency model's design results

Regarding network selection, the PET-VIT network outperforms Unet-based networks in terms of NMAE, PSNR, SSIM, and NCC for PET-CM. However, it requires more generation time. Both Unet and MLP-Unet show comparable performance to PET-VIT in terms of PSNR, SSIM, and NCC, suggesting that they could be viable options as the denoising network in PET-CM. Furthermore, the MLP-Unet demonstrates the highest speed in the generation process.

For PET-CM, using two steps yields the best NMAE and SSIM, while five and ten steps result in superior PSNR, SSIM, and NCC. The performance of 2, 5, and 10 timesteps is similar, but two steps result in the greatest efficiency.

**Table 3**: Quantitative analysis and generation time (G-Time) for synthetic full-dose images from PET-CM using Unet,

PET-VIT (proposed network), and MLP-Unet as the denoising network. The best-performing network(s) are in bold.

| PET-CM | NMAE (%) (↓) | PSNR (dB) (↑) | SSIM (↑) | NCC (↑) | G-time (Sec) |
|---|---|---|---|---|---|
| Unet | 1.428±0.174 | 33.092±0.837 | 0.960±0.010 | 0.965±0.017 | 84.140 |
| MLP-Unet | 1.301±0.130 | 33.628±0.792 | **0.964±0.010** | 0.966±0.018 | 51.351 |
| PET-VIT (proposed) | **1.278±0.122** | **33.783±0.824** | **0.964±0.009** | **0.968±0.011** | 62.321 |

**Table 4**: Quantitative analysis and generation time (G-Time) for synthetic full-dose images from PET-CM with varying timesteps. The best-performing network(s) are in bold.

| PET-CM | NMAE (%) (↓) | PSNR (dB) (↑) | SSIM (↑) | NCC (↑) | G-time (Sec) |
|---|---|---|---|---|---|
| 1 step | 1.876±0.382 | 30.728±0.892 | 0.942±0.016 | 0.943±0.017 | 32.142 |
| 2 steps (proposed) | 1.278±0.122 | 33.783±0.824 | 0.964±0.009 | **0.968±0.011** | 62.321 |
| 5 steps | **1.269±0.164** | **33.811±0.783** | **0.965±0.008** | **0.968±0.011** | 160.761 |
| 10 steps | 1.270±0.134 | 33.790±0.765 | **0.965±0.008** | **0.968±0.011** | 321.481 |

## VI. Discussion

In our study, we introduce the highly efficient PET-CM, based on diffusion probabilistic model framework and optimized to generate superior full-dose PET images from low-dose counterparts. Although renowned diffusion models like DDPM and IDDPM facilitate exceptional synthesis of full-dose PET images, they suffer from time-consuming processes. Conversely, GAN-based alternatives, while faster, significantly compromise image quality. Our PET-CM proposes a proficient solution, forging a favorable equilibrium between speed and quality. It establishes itself as a notable contender in the synthesis of nearly state-of-the-art image with reduced generation times. The model is structured into forward and reverse processes. The forward process incrementally introduces Gaussian noise to a clean full-dose PET image over designated timesteps, ultimately transforming it into pure Gaussian noise. In contrast, the reverse process leverages a dedicated PET-VIT neural network, utilizing associated low-dose PET inputs to effectively denoise the images, and restoring them to their pristine state. The primary distinction between our PET-CM and the work of DDPM-based PET denoising methods (for example, Gong et al.'s work [38]) lies in the underlying frameworks. Although both methodologies are rooted in diffusion dynamics, they diverge significantly in their learning objectives. Our PET-CM adeptly transforms pure Gaussian noise into a pristine PET image, deviating from the traditional DDPMs, which meticulously refine Gaussian noise into a progressively cleaner PET image. Contrary to DDPM's emphasis on learning the incremental noise reduction across timesteps, PET-CM focuses on achieving consistency in the output amidst varying noise levels in PET images. As, our model significantly enhances efficiency, achieving full-dose PET image generation in merely two steps, in stark contrast to usually the 1000 steps required by DDPMs. This efficiency potentially reduces the generation time by a factor of 500 under similar conditions, marking our primary contribution towards high-efficiency PET applications. Additionally, while there is a minor difference in the network architectures used—our model employs Swin-Unet versus DDPM's commonly use of U-net—the fundamental contrast remains within the operational frameworks of the two models. Significantly, we employ the PET-VIT to learn the reverse procedure, functioning as a consistency method that produces identical results across all possible timesteps, to allow image denoising with fewer intermediary steps (for example, transitioning from timestep T to T/2, followed by zero), thus considerably reducing the necessary time. We also integrate a specially designed loss regularization into the consistency model to minimize the mean absolute error between the genuine full-dose PET image and the output of the consistency function. This innovation marks a pioneering step in applying conditional consistency models to medical image translation tasks, promoting a paradigm shift in patient care by enhancing safety, image quality, contrast, and quantification.

As shown in Table 1 and Table 2, PET-CM methods exhibited remarkable quantitative image-based quality, clinical-based quality, and statistical improvements compared to most competing methods (PET-DDPM, PET-CGAN, and PET-cycleGAN). Accordingly, the PET-CM showcases its capacity to maintain image integrity and accurately replicate structural nuances from the full-dose images, therefore potentially enhancing delineation in radiotherapy treatment planning. The efficacy of the PET-CM methodology transcends its notable denoising precision, as delineated in our study. It underscores the implementation of an efficient diffusion-based generation process integral to PET-CM, thereby circumventing the adversarial training paradigm typical of GAN-based techniques. This deliberate omission engenders a more stable training regimen, effectively surmounting a predominant challenge associated with the deployment of GAN-based frameworks. Contrary to GAN methodologies, which demand extensive time investments for hyperparameter optimization (as evidenced by our experiments, spanning several months) and exhibit vulnerability to destabilization upon the incorporation of new data, PET-CM demonstrates remarkable resilience. As illustrated in Figures 2 and 3, despite considerable experimental efforts to refine the networks, PET-CGAN fails to achieve optimal training, showing artifacts. This underscores the difficulty in identifying the optimal set of hyperparameters for CGAN. Conversely, PET-CM facilitates a straightforward training process, obviating the exhaustive search for network parameters. This characteristic significantly mitigates the impact of data variability, thereby substantially curtailing the duration required for hyperparameter adjustment. This inherent stability against data and hyperparameter changes, substantiated by the observations from the network and consistency process presented in Tables 3 and 4, ascertains that PET-CM can be applied without engaging in an exhaustive search for the optimum hyperparameter. Therefore, the PET-CM could potentially be more practical for clinical usage.

On the other hand, the PET-CM demonstrates superior performance compared to the PET-DDPM operating with 50 timesteps. It is worth noting that PET-DDPM underperforms the GAN-based network on a large scale. This could be attributed to the fact that PET-DDPM can achieve comparable image synthesis to PET-IDDPM and PET-CM when the timestep is large (e.g., 1000 timesteps) [29,31], but may not perform as well in our setting with only 50 timesteps. This limitation is visually evident in the generated images (Figures 2 and 3), where the presence of noise indicates that PET-DDPM cannot fully remove noise to generate high-fidelity images. This further suggests that a limited number of timesteps hinders DDPM's ability to generate high-quality images through the effective elimination of noise. To generate a whole-body full-dose PET image of comparable quality, the PET-DDPM necessitates 1000 timesteps, amounting to about 30000 seconds with our hardware settings, rendering it impractical for our evaluation. Moreover, the PET-CM showcases results on par with the PET-IDDPM but operates at a mere 8% of the PET-IDDPM's generation time per patient. This efficiency is attained by establishing a direct connection between the pure noise $X_T$ and the clean full-dose image $X_0$, thereby avoiding an iterative generation process and necessitating only a few timesteps to create high-quality PET images. This pronounced speed not only translates to a more streamlined workflow but also opens avenues for real-time applications where swift decision-making is pivotal. For instance, it can pave the way for the incorporation of PET images in live monitoring and adjustment of radiation doses during real-time image-guided radiotherapy sessions [39], enhancing the precision and responsiveness of the treatments. Furthermore, the reduced generation time implies that healthcare facilities have the potential to process a larger volume of PET images in a reduced timeframe, thereby optimizing resource allocation and increasing the throughput of radiotherapy procedures.

In the study of the hyperparameters of the network and consistency model, PET-CM, using PET-VIT, achieves the best quantitative performance across all metrics, but with increased generation time. MLP-Unet shows comparable performance to PET-VIT with less generation time, indicating its potential as a denoising network in PET-CM if even higher efficiency is required. Furthermore, the investigation underscored the pivotal role of appropriate timestep selection in attaining desirable outcomes. Remarkably, PET-CM exhibited comparable levels of performance across settings of 2, 5, and 10 timesteps, with the 2-timestep configuration presenting the most optimal quantitative efficiency. Delving into the implications of these results reveals that a PET-CM configuration utilizing a mere 2 timesteps is sufficient for state-of-the-art accuracy of the synthetic full-dose PET images. It was discerned that further augmentations in the generation timestep did not yield significant enhancements in image quality, thus indicating a point of diminishing returns beyond the 2-timestep mark.

In addition, the selection of timesteps [1, 75, 150] is strategically designed to facilitate a progressive reduction in noise levels, transitioning from the highest noise intensity at timestep 150 to an intermediate noise level at timestep 75, and culminating at the lowest noise level at timestep 1. Although our framework accommodates transitions between any timesteps within the range 1 to 150 (for example [1,10,150]), the preference for our transitions, however, emerges as the most logical sequence, even though the choice seems to not strongly influence the final performance (as shown in Appendix. C). Nevertheless, it is imperative to acknowledge that the framework retains flexibility, enabling users to adjust the timesteps according to specific requirements or objectives.

The PET-CM methodology exhibits significant advancements in efficiency and quantitative performance compared to other approaches. One limitation is that the realism of using histogram-reduction to simulate low dose PET has not been well studied as downsampling list-mode data [40]. However, theoretically, this method can still be superior to the method of reducing acquisition time, which may introduce inconsistency in uptake distribution but is still commonly used as a surrogate for low dose PET simulation in studies [6,8]. One the other, hand, it is crucial to note, however, that these advancements are predominantly observed in the sphere of 2D PET image synthesis. Our multiple-folded cross-validation experiments have demonstrated the PET-CM's capacity for generalization in 2D PET image synthesis across a dataset comprising 11,200 axial slices. Additionally, the PET-CM has shown its applicability across various anatomical regions, including the brain, chest, and abdomen, within a whole-body PET dataset. This underscores the method's effectiveness in 2D PET image synthesis and potential generalizability on other PET datasets. However, our practical experiments have not yet achieved the desired outcomes when applying PET-CM models to the more complex task of 3D PET image synthesis, a challenge that is also present with other 3D diffusion-based models. The limitations of diffusion-based models, including PET-CM, in synthesizing 3D images may be attributed to the significantly greater complexity of 3D volumes as opposed to 2D slices. Specifically, the process of reverse diffusion is substantially more challenging in the context of 3D PET image denoising. Moreover, expanding our dataset to include more patient cases for the purpose of exploring the effectiveness of the PET-CM approach in 3D PET image denoising represents a critical direction for future research. This is essential to assess the model's generalizability in the denoising of 3D PET images.

Moreover, despite the PET-CM's superior performance over GAN-based methods at eighth- and quarter-dose PET denoising, there are existing studies that achieve PET denoising at even lower doses. The works of Sanaat et al. [15] and Chen et al. [41] introduce GAN-based methodologies for brain PET denoising at 1/20 and 1/50 low-dose levels, respectively. Additionally, Zhou et al.[42] have developed a cycleGAN that demonstrates effective denoising at a 1/100 low-dose level in the lung region. Exploring the extension of the PET-CM to facilitate ultra-low-dose PET denoising (such as at 1/20, 1/50, and 1/100 levels) across the whole body represents another potential avenue for our future research. Additionally, exploring more advanced network architectures [23,43-51] represents a potential focus for our future research.

## VII. Conclusion

In this work, we have introduced a novel diffusion-based method for denoising and synthesizing full-dose PETs from low-dose PETs: the PET consistency model (PET-CM). The method successfully restores noisy PET images of one eighth and one quarter of the full dose to standard high-quality PET images. PET-CM offers a negligible tradeoff in image quality for a thirteen-fold improvement in computational efficiency. PET-CM may prove a valuable tool for improving PET imaging by enabling radiation dose reduction without compromising diagnostic accuracy.

**Acknowledgement**

This research is supported in part by the National Institutes of Health under Award Number R56EB033332, R01EB032680, R01CA272991 and P30CA008748.

**Conflict of interest**

The authors have no conflict of interests to disclose.

# SUPPLEMENT MATERIAL

This document contains supplementary information for the manuscript "Full-dose whole-body PET Synthesis from Low-dose PET Using High-efficiency Diffusion Denoising Probabilistic Model: PET consistency model".

**Appendix. A Network detail of the PET-VIT**

A.1 The network description

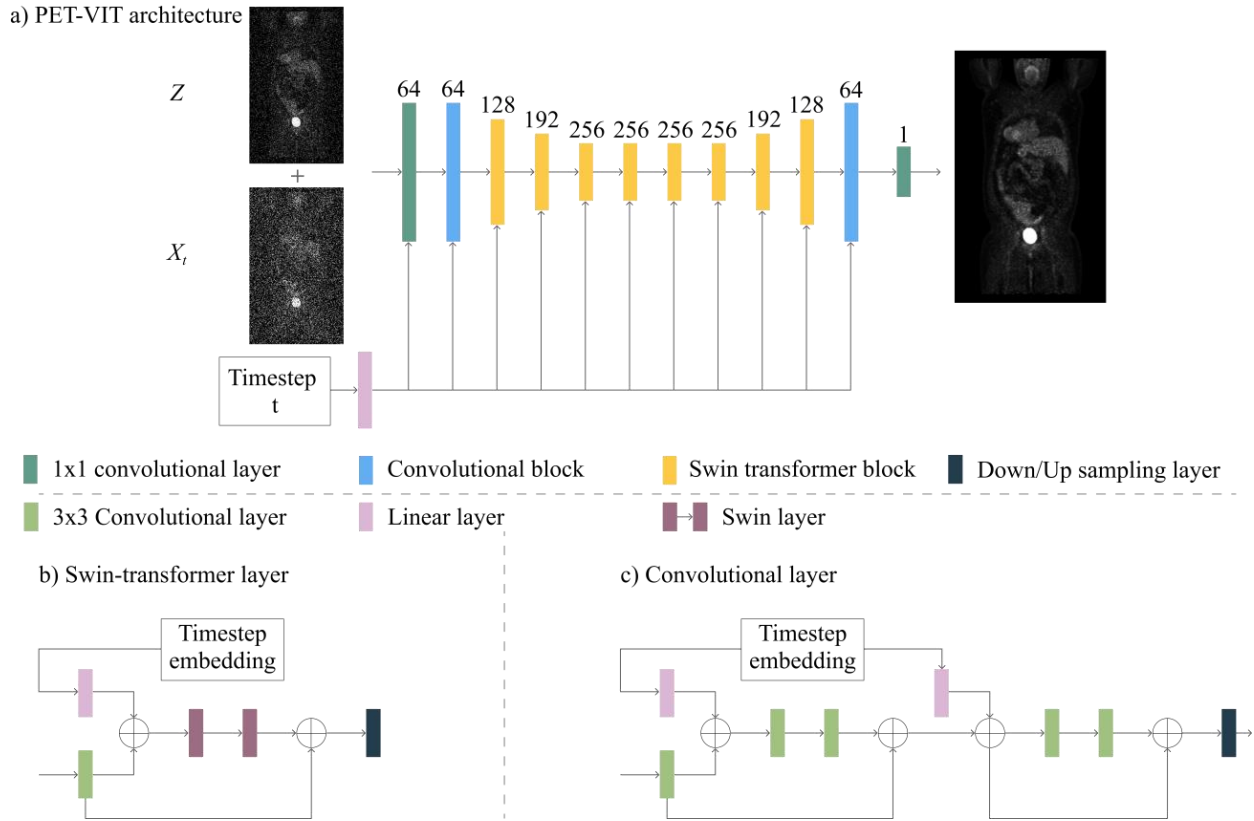
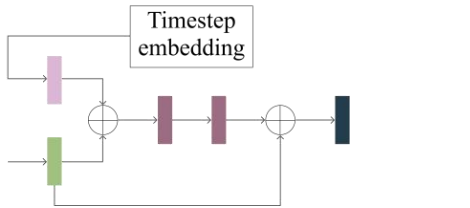
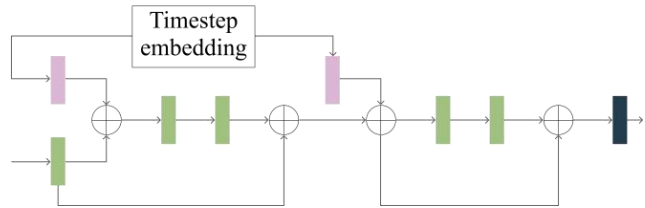

**Figure A. 1**: Visualization of the PET-VIT network using in the PET-CM.

The complete architecture is depicted in Fig A.1. a). Starting with the encoder, the input undergoes a convolutional layer with a 1×1 kernel size and stride of 1 to capture early features. These features are then down-sampled by a factor of 2 in each block. In the encoder architecture, we propose one initial down-sampled convolutional block (as shown in Fig. A.1 c)) to learn local characteristics from inputs with higher resolutions. Subsequently, we incorporate three sequential down-sampled Swin-transformer blocks (as shown in Fig. A.1 b)) to capture global information from lower-resolution features. Following this, two middle Swin-transformer blocks (without down-sampling or up-sampling) are connected to further extract global characteristics. The decoder consists of three up-sampled Swin-transformer blocks and one final up-sampled convolutional block to obtain the features with the original resolution. Eventually, the resulting features are fed into one convolutional layer to estimate the consistency function. On the other hand, the timesteps $t$ are encoded using sinusoidal embedding (SE) [21], where the maximum period of SE is set as 10^6, and the feature dimension is set as 128. These timestep embeddings are then input into all blocks, after adaptive group normalization [21], so the network denoise the input PET according to the corresponding timestep.

Several techniques are additionally applied to stabilize the network training. Firstly, we utilize the "Residual connection" [52] (as shown in Fig. A.1.c)) in each convolutional and Swin-SA block to enhance network stability. Furthermore, we employ "shortcut connections" across the blocks (as shown in Fig. A.1.b)), connecting each encoder block to a decoder block at the same resolution level. This facilitates the transmission of high-resolution information from the encoder to the decoder, enhancing estimation accuracy.

**Table A.1:** PET-VIT details: Convolution channel, convolution kernel size are parameters of the convolutional layers of the convolutional/Swin-transformer layers. Down-sample ratio and up-sample ratio are the parameters of the sampling layers. Window size, attention channel, and attention heads are parameters of the Swin transformer layer.

| PET-VIT | Encoder (5 layers) | | | | | Middle layer | | decoder (5 layers) | | | | |
|---|---|---|---|---|---|---|---|---|---|---|---|---|
| Convolution channel | 64 | 64 | 128 | 192 | 256 | 256 | 256 | 256 | 192 | 128 | 64 | 1 |
| Convolution kernel size | (1, 1) | (3, 3) | (3, 3) | (3, 3) | (3, 3) | (3, 3) | (3, 3) | (3, 3) | (3, 3) | (3, 3) | (3, 3) | (1, 1) |
| Down-sample ratio | (1, 1) | (2, 2) | (2, 2) | (2, 2) | (2, 2) | N/A | N/A | N/A | N/A | N/A | N/A | N/A |
| Up-sample ratio | N/A | N/A | N/A | N/A | N/A | N/A | N/A | (2, 2) | (2, 2) | (2, 2) | (2, 2) | (1, 1) |
| Window size | N/A | N/A | (4, 4) | (8, 8) | (8, 8) | (8, 8) | (8, 8) | (8, 8) | (8, 8) | (4, 4) | N/A | N/A |
| Attention channel | N/A | N/A | 1024 | 1024 | 1024 | 1024 | 1024 | 1024 | 1024 | 1024 | N/A | N/A |
| Attention heads | N/A | N/A | 12 | 24 | 24 | 24 | 24 | 24 | 24 | 12 | N/A | N/A |

**Appendix. B Implementation details of the Heun solver**

Recall that by setting $f(X_t, t) = 0$ and $g(t) = \sqrt{2t}$, we can have a probability-flow ordinary differential equation (PF-ODE) for the reverse process:

$$\frac{dX_t}{dt} = -tS_\theta(X_t, t) \quad [1]$$

the PF-ODE can be solved by neural network or any other ODE solvers. For example, using Heun's second order ODE solver [53], we can obtain the less noisy PET $X_{t-\Delta t}$ (the continuous version of $X_{t-1}$):

$$d = \frac{X_t - X_0}{t} \quad [2]$$

$$d' = \frac{X_t - d\Delta t - X_0}{t - \Delta t} \quad [3]$$

$$X_{t-\Delta t} = X_t - \frac{d+d'}{2}\Delta t \quad [4]$$